\documentclass[useAMS,usenatbib]{mn2e}

\usepackage{defs}
\usepackage{psfig}

\title[Domination of Cold Galaxies across $0<z< 1$]{The Comoving Infrared Luminosity Density: Domination of Cold Galaxies across $0<z<1$}
\author[N. Seymour et al.]{N. Seymour$^{1}$\thanks{E-mail: nps@mssl.ucl.ac.uk}, M. Symeonidis$^{1}$, M.J. Page$^{1}$, M. Huynh$^{2}$, T. Dwelly$^{3}$,   I.M. M$^{\rm c}$Hardy$^{3}$ \and\& G. Rieke$^{4}$\\ 
$^{1}$Mullard Space Science Laboratory, UCL, Holmbury St Mary, Dorking, Surrey, RH5 6NT.\\
$^{2}$Infrared Processing and Analysis Center, MS220-6, California Institute of Technology, Pasadena, CA, 91125, USA.\\
$^{3}$School of Physics \& Astronomy, University of Southampton, Highfield, Southampton, SO17 1BJ.\\
$^{4}$Steward Observatory, Tucson, 85721, USA.}

\begin{document}

\date{ACCEPTED $20^{\rm th}$ November 2009}

\pagerange{\pageref{firstpage}--\pageref{lastpage}} \pubyear{2002}

\maketitle

\label{firstpage}

\begin{abstract}

In this paper we examine the 
contribution of galaxies with different infrared (IR) spectral energy 
distributions (SEDs) to the comoving infrared luminosity density, 
a proxy for the comoving star formation rate (SFR) density. 
We characterise galaxies as 
having either a {\em cold} or {\em hot} IR SED depending upon whether the 
rest-frame wavelength of their peak IR energy output is above or below 
$90\,\um$. Our work is based on a far-IR selected sample both in the 
local Universe and at high redshift, the former consisting of 
{\it IRAS} $60\,\um$-selected galaxies at $z< 0.07$ and the latter 
of {\it Spitzer} $70\,\um$ selected galaxies across $0.1< z\le 1$. 
We find that the total IR luminosity densities for each redshift/luminosity 
bin agree well with results derived from other deep mid/far-IR surveys. 
At $z<0.07$ we observe the previously known results: that moderate luminosity 
galaxies ($L_{IR}<10^{11}\,L_\odot$)
dominate the total luminosity density and that the fraction of cold galaxies
decreases with increasing luminosity, becoming negligible at the highest 
luminosities. Conversely, above $z=0.1$ we find that luminous IR galaxies 
($L_{IR}>10^{11}\,L_\odot$), the majority of which are cold, dominate the IR 
luminosity density. We therefore infer that cold galaxies dominate 
the IR luminosity density across the whole $0< z< 1$ range, hence
appear to be 
the main driver behind the increase in SFR density up to $z\sim1$ whereas
local luminous galaxies are not, on the whole, representative of the high 
redshift population. 
\end{abstract}


\section{Introduction}

The rise in the comoving star formation rate (SFR) density up to $z\sim1$ and 
its subsequent flattening \citep{Lilly:96, Madau:96} has now been well studied 
at several wavelengths 
\citep[e.g.][and references therein]{Bunker:04, Hopkins:06}. 
While the global picture to $z=1$ has been well constrained by 
observation, the details of how or why this change occurs remain largely
unknown. There now
is evidence that star formation depends on both galaxy mass 
\citep[][]{Feulner:05, Juneau:05} and environment 
\citep[][]{Lewis:02,Elbaz:07} presenting a more complicated picture of
galaxy evolution than simple evolution of the luminosity function.

The infrared (IR), and particularly the far-IR, is one of the most powerful 
tracers of star formation as IR luminosity directly scales with SFR 
\citep[][and references therein]{Kennicutt:98b} and far-IR emission
originates from regions of cold dust and gas that constitute the fuel for an 
on-going burst of star formation.
Another advantage of selecting sources at long wavelengths is 
the low contribution by active galactic nuclei (AGN), if present, to the 
total IR luminosity \citep{Alexander:03, Clements:08}.
Studies of the distant Universe at IR wavelengths remain limited and the
most sensitive probe has been surveys with the {\it Spitzer} $24\,\um$ band.
However, this wavelength is relatively far from the peak of all but the 
hottest IR galaxies and progressively shifts to shorter wavelengths at 
higher redshifts where strong spectral features in the observed frame can also 
complicate matters. The few studies done with deep {\it Spitzer} $70\,\um$ 
imaging have found rapid evolution in the total IR luminosity 
function \citep{Huynh:07b, Magnelli:09}. When the IR luminosity 
function is integrated at different redshifts it is found that the 
IR luminosity density increases rapidly up to $z=1$ in a 
similar fashion to the SFR density. Nevertheless, the relative contribution by 
luminosity to the IR luminosity density changes with redshift with 
starbursts ($10\le\log(L_{IR}/L_\odot)< 11$) dominating locally, but with 
luminous IR galaxies (LIRGs: $11\le\log(L_{IR}/L_\odot)< 12$) and ultra-luminous
IR galaxies (ULIRGs: $12\le\log(L_{IR}/L_\odot)< 13$) becoming increasing 
contributors at higher redshifts \citep{LeFloch:05,Magnelli:09}.

In \citet[][hereafter S09]{Symeonidis:09} we studied the IR spectral energy 
distributions (SEDs) of a sample of $70\,\um$ selected galaxies at $z\ge0.1$, 
using a sample of local, $z<0.1$,  {\it IRAS} $60\,\um$ selected galaxies 
for comparison.
We fitted the mid to far-IR photometry of both samples with models from 
\citet{Siebenmorgen:07}. We found that the majority of the $70\,\um$ sources
had IR SEDs which peaked, in $\nu\times F_\nu$, at longer wavelengths than 
galaxies from the local sample with similar luminosities. 
In the local sample we observed a shift  of the IR 
SED peak to shorter wavelengths with increasing luminosity 
\citep[as has previously been noted:][]{Sanders:96, Chapman:03a, Rieke:09}, 
whereas the $70\,\um$ sample had a wide range of peak wavelengths the 
distribution of which varied little with luminosity.
The observation that the IR SEDs of luminous, distant galaxies were on average 
different to their local analogues was in contrast to other recent results, 
\citet[e.g.][]{Magnelli:09} who concluded there was no significant change in 
IR SED of luminous galaxies with redshift.
Models of galaxy evolution in the IR have tended to 
assume that the SEDs of high redshift luminous sources follow the luminosity 
trend seen in local sources
\citep[][]{Lagache:05,Pearson:05,LeBorgne:09,RowanRobinson:09}.
However, as shown in S09, the range of IR SEDs for luminous 
galaxies is much wider at high redshifts than seen locally.

In this paper we examine the contribution of galaxies with different IR SEDs 
to the comoving IR luminosity density (IRLD). 
We will compare our results to earlier work \citep{LeFloch:05} by examining 
the contribution to 
the IRLD by luminosity, but this time with a sample selected at 
$70\,\um$ rather than $24\,\um$. This selection enables us to use a more 
robust selection of IR luminous galaxies as $70\,\um$ lies 
closer to the peak of typical IR SEDs. Hence, the total IR luminosity 
can also be estimated more accurately, especially with constraints (mainly
detections) from $160\,\um$ data. 
We examine the IRLD in more detail by focusing on the contribution 
by IR SED type within each bin. By obtaining an estimate of the peak of the 
IR SED in rest-frame $\nu\times F_\nu$, we can characterise these IR bright 
sources as being either {\em cold} or {\em hot} depending on whether the SED 
peaks above or below $90\,\um$ ($32.2\,$K for a blackbody). We choose this 
wavelength for two reasons. 
Firstly, local IR galaxies appear to have a warm IR component 
($\bar{\lambda}\sim 60\,\um$) associated with dust around young star forming 
regions and a cooler ``cirrus'' component ($\lambda\ge\,100\,\um$) 
associated with more extended dust heated by the interstellar radiation field 
\citep{Lonsdale:87}. Secondly, this $90\,\um$ cut also marks a divide between 
the SEDs of most local ULIRGs and those of cold SMGs seen at high redshift, 
most of which would be classified as cold by our definition
\citep{Chapman:05}.
We note that there is a linear relationship between the {\it IRAS}
colours often used to define the IR galaxies as `cold' or `hot' 
\citep[e.g.][]{Chapman:03a} and the peak wavelength used in this work.

We present our far-IR galaxy sample in \S2, describe our determination of the 
comoving IRLD in \S3. We present our results in \S4 and 
discuss them in \S5. Throughout we use the current `concordance' cosmology: 
$\Omega_M = 1 - \Omega_{\Lambda} = 0.3$, $\Omega_0 = 1$, and 
$H_0 = 70\, \kmpspMpc$ \citep{Spergel:03}.

\begin{figure}
\centerline{
\psfig{file=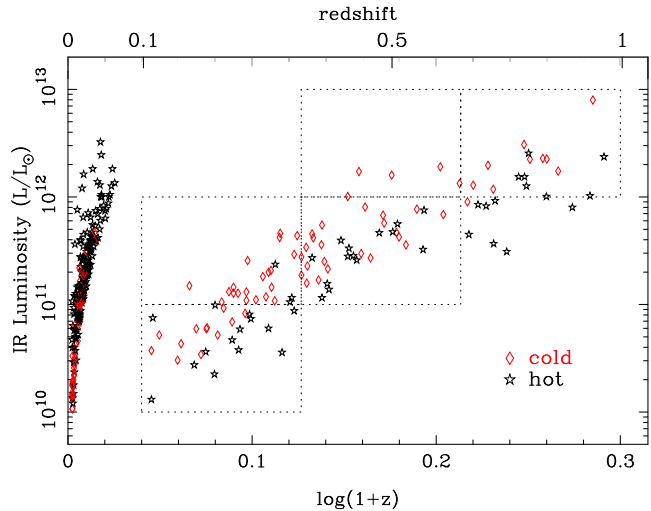,width=8.5cm,angle=270}
}
 \caption{The redshift/total ($8-1000\,\um$) IR luminosity  distribution 
	of the local ($z<0.07$) BGS sample and our {\it Spitzer} $70\,\um$ 
	sample ($z>0.1$) 
	with sources characterised as cold or hot depending on 
	whether their IR SED peaks above or below $90\,\um$ (rest-frame in 
	$\nu\times F_\nu$). The dotted lines indicate the binning chosen to 
	investigate the comoving IRLD by IR luminosity and SED.}
\label{fig:lz}
\end{figure}

\section{The Far-IR Galaxy Sample}

The IR luminous galaxies we use here are taken from the study of S09 who
used a local and distant far-IR selected sample. The local sample
is the {\it IRAS} Bright Galaxy Sample from \citet{Sanders:03} which is 
selected to have $60\,\um$ flux densities above 5.24\,Jy covering the entire 
sky surveyed by {\it IRAS} at Galactic latitudes $|b|>5^\circ$. The higher 
redshift sample we use consists of {\it Spitzer} $70\,\um$ galaxies from 
two well studied extragalactic fields: the Extended Groth Strip 
\citep[EGS covering $\sim0.3^\circ$,][]{Davis:07} 
and the $13^{\rm H}$ {\it XMM-Newton/Chandra Deep Field}
\citep[$13^{\rm H}$ covering $\sim0.3^\circ$,][]{Seymour:08}.

\begin{figure*}
\centerline{
\psfig{file=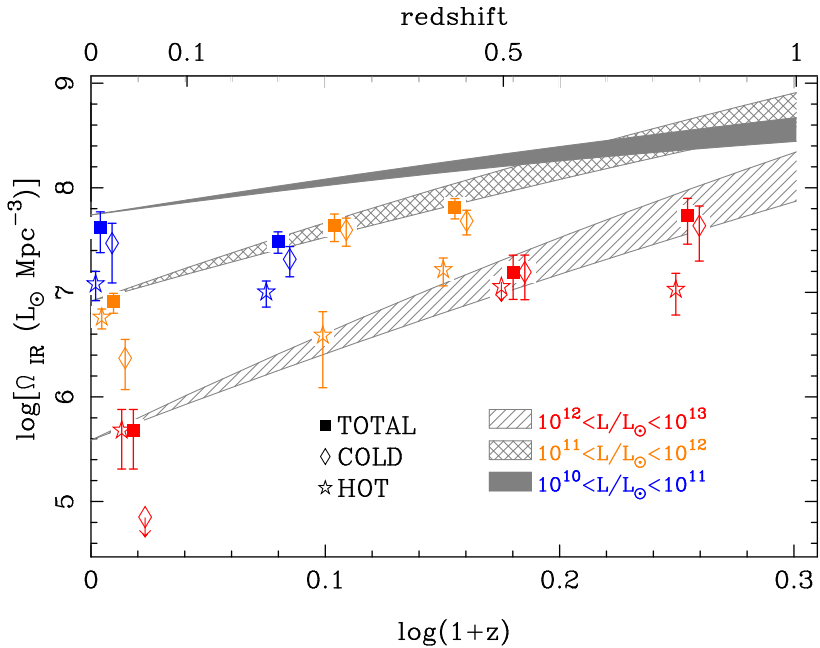,width=10.8cm,angle=270}
}
 \caption{The comoving IR luminosity density (IRLD) plotted as a function of 
	redshift in different luminosity bins: starbursts (blue), LIRGs  (yellow) 
	and ULIRGs (red). For each bin we 
	plot the total IRLD and the separate contribution 
	from the cold and hot 
	galaxies. The hot and cold points are marginally offset in redshift 
	from the total IRLD for clarity. Derivation 
	of the uncertainties and upper limits (representing null detections) 
	is explained in the text.
	Locally ($z$$<$$0.1$) we use the {\it IRAS} Bright Galaxy 
	Sample and at other redshifts we use the {\it Spitzer} $70\,\um$ 
	selected sample from S09. We overlay, for 
	comparison, the IRLD derived from deep $24\,\um$ 
	observations by Le Floc'h et al. (2005, shaded regions). 
	We observe a general trend of hot galaxies dominating the local 
	LIRG and ULIRG bins, but the increase in IRLD at higher 
	redshifts is mainly due to an increase in the contribution from cold 
	galaxies rather than hot galaxies.}

\label{fig:lumden}
\end{figure*}

We restrict our $70\,\um$ sample to sources with $0.1\le z\le1.0$ 
ruling out local resolved sources at $z<0.1$ and very high redshift sources 
where there may not be complete redshift information.
Although the $13^{\rm H}$ field $70\,\um$ sample has complete redshift 
information (40 spectroscopic and 40 photometric) the original EGS 
spectroscopic survey only targetted 
$\sim2/3$ of candidate sources \citep{Davis:07}. 
By examining the original EGS $70\,\um$ sample \citep{Symeonidis:07}
we estimate from broad-band colours  that 101 sources 
lie within $0.1\le z\le1.0$, 66 of which have spectra. Hence, 
in the next section we multiply the area of the EGS by a factor 66/101 to 
correct for those sources not included on the assumption these 66 are 
representative of the parent population.
We impose a further cut of $S_{70\um}>10\,$mJy, to remove lower 
signal-to-noise sources where we are not complete, leaving 
us with 123 sources 
in our $70\,\um$ sample from both surveys. 

We use the total IR luminosities 
and peak wavelengths determined in S09 which were derived
from fits to models from \citet{Siebenmorgen:07}. These models were fit to 
photometry at 12, 25, 60 and $100\,\um$ for the local sample and at 
8, 24, 70 and $160\,\um$ for the {\it Spitzer} $70\,\um$ sample.
The IR luminosity-redshift distribution for both samples
is plotted in Fig.~\ref{fig:lz} where 
we have characterised each source as having a cold or hot IR SED depending 
whether it peaks above or below our fiducial wavelength of $90\,\um$. 

\section{Method}

We determine the IR luminosity density using a version of the classic 
$1/V_{max}$ method \citep{Schmidt:68, Felten:76}. We use 
the following formula to determine the luminosity density in each 
redshift/luminosity bin for the $70\,\um$ sample
\begin{equation}
\Phi_{\rm IR}(L, z)~=~\Sigma_i~ (L_i/V_i)
\end{equation}
where $L_i$ is the total $8-1000\,\um$ IR luminosity (in solar units) of 
a source and $V_i$ is the maximum volume (in Mpc$^3$) a source could occupy 
given the size of the redshift bins, the area of the survey, our $10\,$mJy 
completeness limit and the best fit SED as determined in S09.
The same method was used for the local BGS sample, also using the best 
fit SED found in S09, but the $V_i$ was limited by the $5.24\,$Jy detection 
limit at $60\,\um$ and a redshift range of $0.005\le z\le 0.07$.
Uncertainties are determined using the expression:
\begin{equation}
\delta\Phi_{\rm IR}(L, z)~=~\sqrt{\Sigma_i~(L_i/V_i)^2 + \delta\Phi(\delta L_i)^2}
\label{eq:unc}
\end{equation}
where $V_i$ and $L_i$ are defined as before and the $\delta\Phi(\delta L_i)$ 
term represents the contribution of the uncertainty in the luminosity of 
individual sources within that bin (due to photometric 
redshifts and the modeling in S09).
The binning (see Fig.~\ref{fig:lz}) was defined by the definitions of the 
luminosity class (ULIRG etc.) and the desire to have bins of equal size in 
$\log(1+z)$ space, i.e. approximately equal epochs in cosmic time. We do not 
include a bin to cover the highest redshift LIRGs as our survey is much less 
sensitive 
to cold galaxies in this region of parameter space and hence we are unable 
to derive a representative ratio of hot and cold galaxies. For each 
luminosity-redshift bin we calculated the total comoving IR luminosity 
density as well as that for the cold and hot galaxies separately. 
The redshift used when plotting each bin is the 
mean redshift of all the sources contributing to that bin in $\log(1+z)$ space.
Upper limits for a bin with no cold or hot sources were determined 
from the $97.72\%$ confidence limit (equivalent to $2\,\sigma$) for a null
detection \citep{Gehrels:86} using the volume of the whole bin and the mean 
luminosity of all the galaxies in the bin.

In S09 we discussed how the construction of our sample might affect the 
results. The selection of the $70\,\um$ sample at rest-frame $35-60\,\um$ 
means our selection could potentially favour galaxies that peak 
around those wavelengths, i.e. hot galaxies by our definition, 
contrary to the increasing fraction of cold galaxies. We also 
established that the 
uncertainty in the long-wavelength {\it Spitzer}/$160\,\um$ photometry did not
have a significant effect on the peak wavelength determined from the SED 
fitting. The uncertainties on IR luminosity from the SED fitting, when carried 
over to this work, contribute little to the total uncertainties 
on individual data points in Fig.~\ref{fig:lumden}, i.e. small number 
statistics dominate.

\section{Results}

We present the observed comoving IR luminosity density of our sample in 
Fig.~\ref{fig:lumden} as a function of luminosity and redshift. 
The total IR luminosity densities 
at each luminosity-redshift bin are in good agreement with those from Le 
Floc'h et al. (2005) notwithstanding the potential sample variance 
between small volumes and the different IR SED models used. 
The results
presented here also broadly agree with those of Magnelli et 
al. (2009) who base their work on stacking $24\,\um$ selected sources
at longer wavelengths. The largest disagreement with previous work is at 
$z\sim0.2$ where the 
total IR luminosity density of the starbursts presented here is
about $0.4\,$dex below the value of Le Floc'h et al. (2005). 

\begin{figure}
\centerline{
\psfig{file=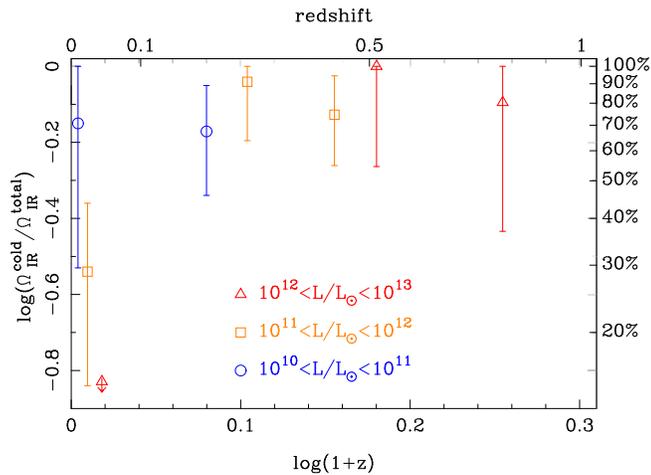,width=8.5cm,angle=270}
}
\caption{The fraction of the comoving IR luminosity density for different 
redshift/luminosity bins due to cold galaxies ($\lambda_{\rm peak}>90\,\um$)
also given as a percentage on the right axis. 
Note by this definition we have no cold galaxies in our local ULIRG sample. }
\label{fig:frac}
\end{figure}

Examining the contribution by IR SED to each luminosity-redshift bin,
we find that there is significant variation in the relative contribution 
of cold and hot galaxies with luminosity and redshift. Locally we see cold 
sources dominating the starburst bin, but at 
higher luminosities they contribute less; cold galaxies comprise 
only one third of the LIRG bin and they contribute nothing to the ULIRG bin. 
The contribution of cold galaxies to the IR luminosity 
density from starbursts remains fairly constant at $60-70\%$ over
the redshift range we can detect them, $0\le z\le0.2$. 
However, for both LIRG and 
ULIRG bins the contribution from cold galaxies increases considerably 
at higher redshifts. Cold galaxies become the dominant contributor to the IR 
luminosity density in these bins above $z=0.1$, in contrast to the 
situation seen in the local Universe. In fact, cold galaxies are responsible 
for most of the rise with redshift of the IR 
luminosity density in both the LIRG and ULIRG bins.

In Fig.~\ref{fig:frac} we plot the fraction of the total IR luminosity 
density in each redshift/luminosity bin 
due to cold galaxies. We can see how the contribution of cold galaxies to 
the starburst luminosity density remains constant up to $z\sim0.2$, 
although our data here cannot constrain that fraction at higher redshifts.
The contribution of cold galaxies to the LIRG and ULIRG luminosity density
increases dramatically with redshift; locally the contribution is very low
or negligible, but at higher redshift cold galaxies dominate.
We find that our results do not change qualitatively if we change our choice 
of fiducial wavelength (that divides between hot or cold galaxies) by $\pm10\,\um$.

\section{Discussion}

In S09 we demonstrated that the IR SEDs of individual LIRGs and ULIRGs at 
$0.1\le z\le 1.0$ peak at longer wavelengths, and span a wider range, than 
local galaxies at similar luminosities. 
Other changes in the properties of star 
forming galaxies have occured since $z\sim1$. Star formation shifts from
massive galaxies at high redshifts to progressively less massive 
galaxies at lower redshifts \citep[e.g.][]{Cowie:96,Juneau:05}.
A shift is also seen in environment of the most strongly star forming 
galaxies with star formation tending to occur in over-dense environments 
at $z\sim 1$ \citep{Elbaz:07}, a reverse of the trend seen locally.
While the dependence of star formation on stellar mass and environment inform 
us about the star formation history and potential or current galaxy mergers, 
information about the IR SED is a direct measure of the physical conditions 
in galaxies underlying major episodes of star formation.

In this paper we have expanded the results of S09 by deriving IR 
luminosity densities at different luminosities and epochs. We observe that 
the total IR luminosity densities derived from a $70\,\um$ sample
are broadly consistent with results derived from shorter wavelengths 
(Fig~\ref{fig:lumden}). The low IR luminosity density of the $z\sim0.2$ 
starburst bin compared to other results is likely due to the effect of 
sample variance over the small cosmic volume probed.
We also separate our sample into galaxies that can be characterised as having 
cold or hot SEDs and study their contribution to each redshift/luminosity bin.
We find two striking new results. Firstly, we observed a rapid change in the 
make up of LIRG and ULIRG bins: locally they are dominated by hot galaxies, 
but at $z>0.2$ they are dominated by cold galaxies. Secondly, cold galaxies
appear to be the dominant contributor to the total IR luminosity density
over $0\le z\le 1.0$; when starburst galaxies dominate at $z<0.2$ we find that
they are mainly cold and by the time LIRGs and ULIRGs begin to dominate at 
higher redshifts they are also mainly cold.
Hence, a cold mode of star formation
has dominated galaxy evolution since $z\sim1$ and is 
likely responsible for the increase in the star formation rate density up 
to $z=1$. 

While Fig.~\ref{fig:lumden} does dramatically show the rise of cold galaxies 
in the LIRG and ULIRG bins at higher redshifts we note that hot galaxies 
are still found at those redshifts. In fact the results presented here do not 
rule out a rise with redshift in the number (and luminosity density) of hot 
galaxies. Therefore, the distribution of types of SEDs remains quite wide
at high redshift.
Fig.~\ref{fig:frac} most strikingly demonstrates the evolution of the 
relative contribution of cold galaxies to each redshift/luminosity bin. 
The rise in the contribution of cold galaxies to the LIRG and ULIRG bins
is most rapid between $z\sim0$ and $z\sim0.3$. 
We now consider if AGN could be responsible for the change in the fraction 
of cold galaxies. As a powerful AGN  would make the IR SED hotter the 
presence of an AGN  
can not explain the prevalence of cold galaxies at high redshift. AGN 
could, however, be the cause of the high fraction of hot galaxies locally.
Studies have shown that the incidence of AGN are low in local LIRGs, 
$10-20\%$, (Petric et al., 2009, in press), but while it is higher in 
ULRIGs, $\sim40\%$ \citep{Farrah:07} the $8-1000\,\mu$m 
IR luminosity remains dominated by 
star formation. We note that at higher redshift the incidence of AGN, 
and their contribution to total 
IR luminosity, also remains low \citep{Alexander:03, Clements:08}.
Hence, AGN are not a significant contributor to the IR luminosity
of the sources studied here and therefore are not responsible for any of 
the trends with redshift.

The prominence of cold galaxies discovered here might suggest that there 
could be a substantial contribution to the IR emission of luminous galaxies 
at high redshift from a cold cirrus component akin to that observed in local 
IR galaxies.
Such an interpretation is in agreement with the 
observation that many of the distant, cold, SMGs are extended on scales of 
$\sim10\,$kpc \citep{Chapman:04} in contrast with the compact ULIRGs seen
locally, and is also consistent with 
the change in the mean IR SED of luminous galaxies as inferred from the 
observed $70\,\um$ to radio flux density ratio of high redshift star forming 
galaxies \citep{Seymour:09}.
In addition, changes 
in dust properties (such as opacity, grain distribution etc.) with redshift 
are also a possible cause of the larger spread in SED types at high redshift. 

Most galaxy evolution models 
\citep[e.g.][]{Lagache:05,Pearson:05,LeBorgne:09,RowanRobinson:09} 
assume that the IR SED is only dependent on IR luminosity, i.e. they typically
use certain templates for star forming galaxies of a given luminosity, often 
using local galaxies like Arp 220 (which can be characterised as having a hot
IR SED by our definition) for the most luminous population. This 
choice was often necessary as there were few constraints 
on any change with redshift until now. 
Models based on cooler SEDs could , for example, reconcile the flat or 
decreasing star formation rate density above $z=1$ and the far-IR and 
sub-millimetre source counts.

In conclusion, we have found that the star formation history of the Universe
is dominated by cold galaxies across $0<z<1$ and that the hot luminous 
IR galaxies we see locally are not representative of the luminous galaxies
that make up the bulk of the star formation at $0.1<z<1$.

\section*{Acknowledgments}

We thank Andrew Hopkins for useful discussions.
This work is based in part on observations made with the {\it Spitzer Space 
Telescope}, which is operated by the Jet Propulsion Laboratory, California 
Institute of Technology under a contract with NASA. Support for this work 
was provided by NASA through an award issued by JPL/Caltech.  The National 
Radio Astronomy Observatory is a facility of the National Science Foundation 
operated under cooperative agreement by Associated Universities, Inc. 
This work was partially supported by JPL/Caltech contract 1255094 to the 
University of Arizona.


\bsp

\label{lastpage}

\end{document}